\documentclass[reprint,final,floatfix,nofootinbib,superscriptaddress,longbibliography]{revtex4-2}

\usepackage{lmodern,mathtools,amssymb,mathrsfs,bbm,microtype,bm} 
\usepackage{color,graphicx} 
\definecolor{bluesmoke}{rgb}{0.207843,0.415686,0.623529}
\usepackage[
  allcolors=bluesmoke,
  colorlinks,
  pdftex,
  pdftitle={Phase transitions and microphases in elastomers. II. Anisotropy-driven morphologies},
  pdfauthor={Manu Mannattil, Second Author, and Third Author},
  pdfkeywords={Phase separation, pattern formation, elasticity, elastomers, phase-field modeling},
  pdfsubject={Statistical Physics, Polymers \& Soft Matter, Condensed Matter \& Materials Physics},
  hypertexnames=false
]{hyperref}
\DeclareMathOperator{\trace}{tr}

\def\abs#1{\ensuremath{\left|#1\right|}}

\def\dd{\ensuremath{\mathrm{d}}}
\def\div{\ensuremath{\boldsymbol{\nabla\cdot}}}

\def\grad{\ensuremath{\boldsymbol{\nabla}}}
\def\gradX{\ensuremath{\boldsymbol{\nabla}_{\!\!\bm{X}}}}
\def\kbt{\ensuremath{k_{\text{B}}}T}
\def\mz{\ensuremath{m_{0}}}
\def\ms{\ensuremath{m_{\text{s}}}}
\def\NN{\ensuremath{\nonumber}}
\def\phic{\ensuremath{\phi_{\text{c}}}}
\def\phiz{\ensuremath{\phi_{0}}}
\def\psiz{\ensuremath{\psi_{0}}}
\def\qm{\ensuremath{q_{\text{m}}}}
\def\bqm{\ensuremath{\bm{q}_{\text{m}}}}
\def\strain{\ensuremath{\bm{\varepsilon}}}
\def\stress{\ensuremath{\bm{\sigma}}}
\def\Tc{\ensuremath{T_{\text{c}}}}
\def\Tm{\ensuremath{T_{\text{m}}}}
\def\trans{\ensuremath{^{\mathsf{T}}}}
\def\qz{\ensuremath{\hat{q}_{z}^{2}}}
\def\gammaz{\ensuremath{\gamma_{0}}}
\def\unit{{\kern2.333pt}}

\definecolor{hlcolor}{rgb}{0.646,0.165,0.165}

\makeatletter
 \def\bibsection{%
  \par
  \begingroup
  \baselineskip26\p@
  \bib@device{\hsize}{72\p@}%
  \endgroup
  \nobreak\@nobreaktrue
  \addvspace{19\p@}%
}%
\makeatother

\begin{document}

\title{Phase transitions and microphases in elastomers. II.~Anisotropy-driven morphologies}
\author{Manu Mannattil}
\email{manu.mannattil@posteo.net}
\thanks{Present address: School of Engineering and Applied Sciences, Harvard University, Cambridge, Massachusetts 02138, USA.}
\affiliation{School of Chemistry, and Center for Physics and Chemistry of Living Systems, Tel Aviv University, Tel Aviv 69978, Israel}
\affiliation{School of Physics and Astronomy, and Center for Physics and Chemistry of Living Systems, Tel Aviv University, Tel Aviv 69978, Israel}
\author{David Andelman}
\email{andelman@tauex.tau.ac.il}
\affiliation{School of Physics and Astronomy, and Center for Physics and Chemistry of Living Systems, Tel Aviv University, Tel Aviv 69978, Israel}
\author{Haim Diamant}
\email{hdiamant@tauex.tau.ac.il}
\affiliation{School of Chemistry, and Center for Physics and Chemistry of Living Systems, Tel Aviv University, Tel Aviv 69978, Israel}

\begin{abstract}
  In swollen elastomers, elasticity can arrest macroscopic demixing between the polymer network and the solvent, producing stable domains of finite size.
  While previous theories have focused on isotropically swollen elastomers, anisotropy strongly influences the phase behavior of many related systems, such as gels and crosslinked polymer blends.
  Building on the approach developed in Part I of this two-part paper, we investigate the effects of two experimentally induced sources of anisotropy: uniaxial swelling and stiffness gradients.
  We show that uniaxial swelling can lead to the formation of stable lamellar microphases and modify phase behavior, while stiffness gradients can cause spatial variations in the characteristic microphase size.
\end{abstract}
\maketitle

\section{Introduction}
\label{sec:introduction}

Anisotropy can profoundly influence phase behavior and pattern formation across a wide range of physical systems.
Some examples include the nematic--isotropic transition in liquid crystals in the presence of external fields~\cite{gennes1993}, crystallographically directed spinodal decomposition in cubic crystals~\cite{cahn1962}, chessboard-like patterns phase-separating alloys driven by anisotropic elastic interactions~\cite{le-bouar1998}, and the reorientation and stabilization of block-copolymer microphases under shear~\cite{hamley2000} and electric fields~\cite{amundson1993}.

Elastic microphase separation (EMPS) is a form of phase separation observed in swollen elastomers, where thermodynamic demixing between the elastomeric polymer network and solvent is arrested by network elasticity, leading to stable solvent-rich domains of finite size (microphases)~\cite{fernandez-rico2024}.
EMPS is similar in spirit to network--solvent phase separation observed in certain polyacrylamide gels upon cooling~\cite{tanaka1977,tanaka1978,li1989,onuki1993}.
Although microphases are not formed in these gels, as in many phase-separating systems, anisotropy can have drastic effects on the phase behavior.
For example, in uniaxially swollen gels, density fluctuations and domain growth preferentially occur parallel or perpendicular to the stretching direction~\cite{onuki1988}, with similar effects also reported in crosslinked polymer blends~\cite{rouf1994}.

These observations highlight the need for a theoretical framework that can account for anisotropic effects in EMPS.
Previous works on EMPS have focused exclusively on isotropically swollen elastomers, often employing nonlocal linear theories of elasticity.
For a summary of these works \cite{qiang2024,mannattil2025,paulin2026,oudich2026,wang2025,thewes2026,safran2026}, see Part I~\cite{part1}.
However, the elasticity of polymer networks and other rubbery materials is fundamentally nonlinear~\cite{treloar1975}.
Hence, any linear theory must arise as the small-strain limit of a more general, nonlinear one.
Unfortunately, we are not aware of a widely accepted procedure by which a nonlinear theory can be linearized to yield nonlocal linear elasticity,
although some attempts exist~\cite{eringen1987,picu2003}.

In Part~I~\cite{part1}, we showed that EMPS can be understood within the framework of conventional elasticity.
This approach provides a consistent linearization of nonlinear rubber elasticity and enables the study of anisotropic effects in this second part (Part~II).
Motivated by the experiments of Ref.~\cite{fernandez-rico2024}, we examine two sources of anisotropy in EMPS: \emph{i})~uniaxial swelling and \emph{ii})~stiffness gradients.
We show that uniaxial swelling can stabilize lamellar morphologies whose orientation depends on the direction and magnitude of the applied stretch or compression, while stiffness gradients lead to spatial variations in the characteristic microphase domain size.

This paper is organized as follows.
In Sec.~\ref{sec:model}, we describe our approach in broad strokes, with more detailed discussions in Appendices~\ref{app:continuum} and \ref{app:numerical}.
Section \ref{sec:uniaxial} discusses the general phase behavior of an elastomer that is isotropically and uniaxially swollen.
Experimental comparison is summarized in Sec.~\ref{sec:experiments}.
Finally, we conclude in Sec.~\ref{sec:conclusion}.

\section{Model}
\label{sec:model}

As in Part I~\cite{part1}, we consider swollen elastomers composed of a charge-neutral, crosslinked polymer network and a solvent.
We do not study the swelling dynamics but assume that the elastomer has already reached thermodynamic equilibrium and is in an affinely deformed state.
Unlike in Part I, however, the swelling is not assumed to be isotropic, and general affine deformations are allowed.
Such anisotropic swelling may arise from externally applied stresses or from the geometry of the domain confining the elastomer.
Regardless of its origin, affine swelling implies that the network volume fraction $\phiz$ is spatially uniform in the equilibrium state.

Upon lowering the temperature $T$, the network volume fraction $\phi(\bm{x})$ at a point $\bm{x} = (x, y, z)$ deviates from $\phiz$.
Here $\bm{x}$ is an Eulerian coordinate describing the deformed elastomer.
As in the isotropic case, the total free energy $\mathscr{F}$ of the swollen elastomer is expressed as a sum of thermodynamic and elastic contributions, and is given by
\begin{equation}
  \mathscr{F}[\psi, \bm{u}] = \mathscr{F}_{\text{GL}}[\psi] + \mathscr{F}_{\text{el}}[\bm{u}],
  \label{eq:free_total}
\end{equation}
where the order parameter is $\psi(\bm{x}) = \phi(\bm{x}) - \phic$ and $\phi_c$ is the critical volume fraction.
Throughout this work, we assume that the network--solvent system is close to a critical point $(\phic, \Tc)$ and take the Ginzburg--Landau free energy $\mathscr{F}_{\text{GL}}$ in the usual form
\begin{align}
  \mathscr{F}_{\text{GL}}[\psi] & = \int \dd^{3}x\,\bigg[\frac{1}{2}a(T-\Tc)\psi^{2} + \frac{1}{4}b\psi^{4}\NN \\
                                & \quad+ \frac{1}{2}\kappa\abs{\grad\psi}^{2} + \eta(\psi - \psiz)\bigg].
  \label{eq:free_GL}
\end{align}
In the above free-energy expansion, the parameters $a, b > 0$ are phenomenological, the gradient squared term with $\kappa > 0$ penalizes compositional inhomogeneities, and $\eta$ is the chemical potential that acts as a Lagrange multiplier to set the spatial mean of $\psi(\bm{x})$ to $\psiz = \phiz - \phic$.

The elastic energy $\mathscr{F}_{\text{el}}[\bm{u}]$ appearing in Eq.~\eqref{eq:free_total} is the linearized elastic energy associated
with the network deformation. We describe it using a displacement field $\bm{u}(\bm{x})$
written as a function of the Eulerian coordinates $\bm{x}$ of the deformed state.
In Part I~\cite{part1}, $\mathscr{F}_{\text{el}}$ was constructed using the elastic moduli of an isotropically swollen elastomer,
which are well-known and obtained by linearizing standard nonlinear elastic theories~\cite{onuki1993}.
However, consideration of general affine deformations requires a derivation starting from the full nonlinear elastic energy as we do below.

The elastic energy of nonlinear, rubbery materials is typically expressed in terms of the coordinates describing the material reference state~\cite{reddy2013}.
To linearize such a theory, we first describe all deformations from the dry, unswollen state of the elastomer, which is characterized by the three-dimensional (3D) material coordinates $\bm{X}$.
The polymer network in the elastomer undergoes two successive deformations before reaching its final, equilibrium configuration.
First, it swells by absorbing the solvent, and then it undergoes a second deformation during any potential phase separation.

We can describe the final equilibrium configuration using the 3D spatial coordinates
\begin{equation}
  \bm{x} = \mathsf{A}\bm{X} + \bm{u}(\bm{X}),
  \label{eq:deformation}
\end{equation}
where $\mathsf{A}$ is a $3\times3$ matrix with constant entries describing the affine deformation during swelling.
The displacement field $\bm{u}(\bm{X})$, measured relative to the affinely swollen state, captures deformations that occur during phase separation.
The deformation gradient tensor associated with Eq.~\eqref{eq:deformation} is
\begin{equation}
  \mathsf{F} = \gradX\bm{x}(\bm{X}) = \mathsf{A} + \gradX\bm{u}(\bm{X}),
  \label{eq:defgrad}
\end{equation}
with $\gradX(\cdot)$ representing the gradient with respect to $\bm{X}$.

The total, nonlinear elastic energy of an elastomer is
\begin{equation}
  \mathscr{F}_{\text{el}} = \frac{1}{2}\int\dd^{3}X\, W(\mathsf{F}),
  \label{eq:elastic_energy}
\end{equation}
where the strain-energy density $W$ is usually taken to be a function of the deformation gradient tensor $\mathsf{F}$ or its invariants,
which themselves are functions of the coordinates $\bm{X}$.
The exact form of $W$ depends on the material under consideration, and often requires the usage of sophisticated theories~\cite{han1999}.
In simpler, classical theories, $W$ is taken to be proportional to the polymer strand density $\nu$ of the underlying network.
A polymer strand is a segment of the polymer network that connects two neighboring crosslinks,
and $\nu$ is the mean number of such strands per unit volume.
In Ref.~\cite{fernandez-rico2024}, the Young's modulus of the elastomer was found
to have a monotonic relationship with the crosslinker concentration and, consequently, with the strand density $\nu$.
Hence, for our purposes, classical forms of $W$ are sufficient.

A classical strain-energy density $W$ widely used~\cite{onuki1993,onuki2002,jia2021} to model rubbers is
\begin{equation}
  W(\mathsf{F}) = \frac{1}{2}\nu\kbt(I_{1} - 2\ln{J} - 3), \label{eq:energy_density}
\end{equation}
where $k_{\text{B}}$ is the Boltzmann constant and $J = \det\mathsf{F}$ is the Jacobian determinant of the deformation-gradient tensor $\mathsf{F}$, Eq.~\eqref{eq:defgrad}.
Also, $I_{1} = \trace(\mathsf{B})$ is the trace of the left Cauchy--Green deformation tensor $\mathsf{B} = \mathsf{F}\mathsf{F}\trans$.
The second (logarithmic) term in $W$, originally introduced by Flory~\cite{flory1953}, is required to ensure that the material remains unstressed in the presence of zero strain~\cite{onuki1993}.

We substitute  Eqs.~\eqref{eq:defgrad} and \eqref{eq:energy_density} in Eq.~\eqref{eq:elastic_energy} and expand the elastic energy $\mathscr{F}_{\text{el}}$
to quadratic order in the displacement field $\bm{u}$, written now in terms of the spatial coordinates $\bm{x}$
(see Appendix~\ref{app:continuum} for details).
The total elastic energy in Fourier space then takes the form
\begin{align}
  \mathscr{F}_{\text{el}}[\bm{u}] & = \frac{1}{2}\nu\kbt \phiz\int \frac{\dd^3{q}}{(2\pi)^{3}}\big[(\bm{q}\cdot\bm{u}_{\bm{q}})(\bm{q}\cdot\bm{u}_{-\bm{q}})\NN \\
                                  & \quad+ \mathsf{A}_{jl}\mathsf{A}_{kl}\,q_{j}q_{k}(\bm{u}_{\bm{q}}\cdot\bm{u}_{-\bm{q}})\big] ,
                                  \label{eq:free_el}
\end{align}
where $\bm{u}_{\bm{q}} = \int \dd^{3}x\, \mathrm{e}^{-i\bm{q}\cdot\bm{x}}\, \bm{u}(\bm{x})$ is the Fourier transform of $\bm{u}(\bm{x})$ with $\bm{q} = (q_{x}, q_{y}, q_{z})$.
The repeated indices $j, k, l$ run from 1 to 3, which correspond to the $x, y, z$ directions.
Repeated indices are summed over as usual.
We note that in theories describing gels, similar expressions for elastic energy have been obtained
by other authors using somewhat different arguments~\cite{onuki1993,panyukov1996}.

\subsection*{Nonlocal effects}

Following the approach of Part I~\cite{part1}, the fields $\psi$ and $\bm{u}$ are not independent as they are related through local volume conservation.
Although both describe the same polymer network, they operate on different length scales: $\psi$ captures compositional inhomogeneities at molecular scales, whereas $\bm{u}$ describes elastic deformations at mesoscopic scales.
To account for this difference in length scales, we consider a material conservation relation of the form (Appendix~\ref{app:continuum})
\begin{equation}
  \div\bm{u} \approx -\phic^{-1}\bar{\psi}(\bm{x}).
  \quad
  \label{eq:matcons}
\end{equation}
Here, $\bar{\psi}$ is a continuum field obtained by filtering
$\psi$ so that the short-wavelength compositional fluctuations that are not expected to stress the network are suppressed:
\begin{equation}
  \bar{\psi}(\bm{x}) = \int\dd^{3}{y}\,K(\bm{x} - \bm{y})\,\psi(\bm{y}).
  \label{eq:blur}
\end{equation}
The usage of a coarse-graining kernel $K$ makes the material conservation relation, Eq.~\eqref{eq:matcons}, nonlocal.
Unlike Part I~\cite{part1}, where we assumed $K$ to be isotropic, here we choose it to be a general Gaussian kernel of the form
\begin{equation}
  K(\bm{x}) = \abs{\det (2\pi\mathsf{H})}^{-1/2} \mathrm{e}^{-\frac{1}{2}\bm{x}^\mathsf{T}\mathsf{H}^{-1}\bm{x}}, \label{eq:kernel}
\end{equation}
where the matrix $\mathsf{H}$ is a length-covariance matrix that controls the extent of coarse-graining in various directions.
The entries of $\mathsf{H}$ have dimensions of length squared, and we take $\mathsf{H}$ to be a $3\times 3$ symmetric, positive-definite matrix satisfying $\bm{x}^{\mathsf{T}}\mathsf{H}\bm{x} > 0$ for all $\bm{x}$.

We also assume that phase separation is diffusion dominated and excites only bulk longitudinal modes.
Such an assumption is valid when the system size is much larger than the size of the compositional fluctuations and is standard in the theory of gels~\cite{onuki2002}.
Under this assumption, the elastic energy, Eq.~\eqref{eq:free_el}, can be written entirely in terms of the order parameter $\psi$ using the material-conservation relation, Eq.~\eqref{eq:matcons}~\cite{onuki2002}.
Fourier transforming Eq.~\eqref{eq:matcons} and noting that the Fourier transform of the kernel in Eq.~\eqref{eq:kernel}
is $K_{\bm{q}} = \mathrm{e}^{-\frac{1}{2}\bm{q}\trans\mathsf{H}\,\bm{q}}$, we obtain
\begin{equation}
  i\bm{q}\cdot\bm{u}_{\bm{q}} = -\phic^{-1}\psi_{\bm{q}}\,
  \mathrm{e}^{-\frac{1}{2}\bm{q}\trans\mathsf{H}\,\bm{q}} \, .
\end{equation}
Substituting the above result in Eq.~\eqref{eq:free_el} and keeping only the longitudinal modes,
we find that the total free energy, Eq.~\eqref{eq:free_total}, reduces to
\begin{equation}
  \mathscr{F}[\psi] = \mathscr{F}_{\text{GL}}[\psi] +
  \frac{1}{2}\int\frac{\dd^{3}q}{(2\pi)^{3}}M_{\bm{q}}\psi_{-\bm{q}}\psi_{\bm{q}},
  \label{eq:free_total_simple}
\end{equation}
where $M_{\bm{q}}$ is an effective $\bm{q}$-dependent longitudinal modulus given by
\begin{equation}
  M_{\bm{q}} = \left(\frac{\nu\kbt\phiz}{\phic^{2}}\right)\left(1 + \mathsf{A}_{jl} \mathsf{A}_{kl}\frac{q_{j}q_{k}}{q^{2}}\right)\mathrm{e}^{-\bm{q}\trans\mathsf{H}\,\bm{q}}, \label{eq:M_general}
\end{equation}
with $q = \abs{\bm{q}}$.

\begin{figure*}
  \centering\includegraphics{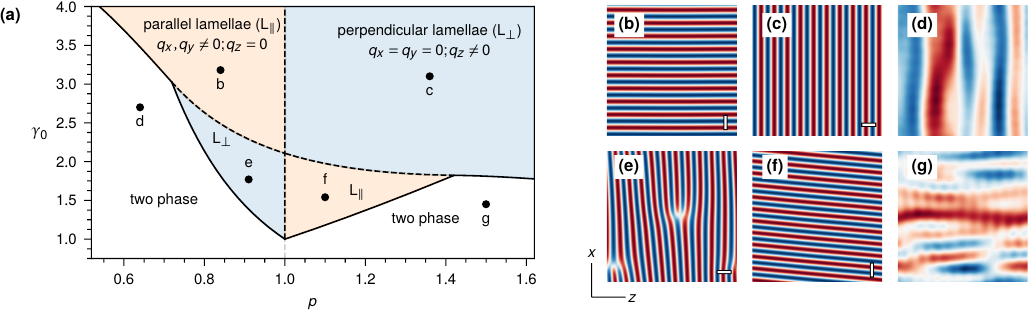}
  \caption{Phase diagram indicating the stability of the various phases of an elastomer that is uniaxially stretched or compressed in the $(p, \gammaz)$ plane.
    These phases are the ones that appear first during a temperature quench at the critical volume fraction where the order parameter $\psiz = 0$.
    The anisotropy factor $p < 1$ denotes compression, $p > 1$ denotes stretching, and $\gammaz$ is the elastocapillary number.
    For illustrative purposes, we set the critical parameters to $(\phic, \Tc) = (0.2, 0)$, and all other free-energy parameters are set to unity.
    The temperature $T$ varies across the diagram, and at a given point, it is taken to be slightly less than the microphase separation temperature $\Tm(0)$ at that point given by Eq.~\eqref{eq:Tm}.
    In the colored regions, lamellar microphases form either along the stretched $z$-axis (L$_\parallel$, $q_{z} = 0$) or perpendicular to it (L$_\perp$, $q_{z} \neq 0$).
    The dashed curves separating these regions are analogous to second-order transition lines and the solid curves to first-order triple lines (where three phases coexist).
    Microphase separation is not possible below $\gammaz = 1$ (Lifshitz point).
    The uncolored regions represent two-phase coexistence after macrophase separation.
    In these regions, during early stages of phase separation, domain growth is fastest perpendicular to the $z$-axis when $p < 1$ and parallel to it when $p > 1$.
    (b)--(g) Representative morphologies of the phase-separated domains, with panels (d) and (g) showing early stages of domain growth in the two-phase regions.
    Polymer-rich and solvent-rich regions are shown in blue and red, respectively.
    The $(p, \gammaz)$ values used to obtain these morphologies have been marked in (a) using black dots.
    The scale bars indicate the domain sizes predicted from the magnitudes of the most unstable wavevectors, Eqs.~\eqref{eq:qm_perp} and \eqref{eq:qm_par}.}
  \label{fig:uniphase}
\end{figure*}

\section{Uniaxial swelling}
\label{sec:uniaxial}

When elastomers undergo uniaxial swelling---either due to an externally applied stress or because their swelling is constrained within a noncubic domain---the resulting equilibrium swollen state differs from that of an isotropically swollen gel.
In such cases, the diagonal entries of the affine deformation matrix $\mathsf{A}$ are no longer equal.
However, as with the isotropic case, its determinant $\det\mathsf{A} = \phiz^{-1}$ is always constrained by volume conservation (see Appendix~\ref{app:continuum}).

Let us consider the situation in which the uniaxial deformation (compression or elongation) due to swelling is along the $z$-axis, and $\bm{x} = (x, y, z)$ is the spatial coordinate describing the swollen elastomer.
In this case, $\mathsf{A}$ remains diagonal, and we takes its entries to be
\begin{equation}
  \mathsf{A}_{xx} =
  \mathsf{A}_{yy} = p^{-1/2}\phiz^{-1/3}
  \quad\text{and}\quad
  \mathsf{A}_{zz} = p\phiz^{-1/3}.
\end{equation}
Clearly, $\det\mathsf{A} = \phiz^{-1}$ is satisfied, and $p > 0$ is a dimensionless anisotropy factor.
It quantifies the degree of anisotropy in the initial swelling, with $p = 1$ corresponding to the isotropic case.
We introduce the same anisotropy as in $\mathsf{A}$ in the length-covariance matrix $\mathsf{H}$, which remains diagonal.
The entries of $\mathsf{H}$ have dimensions of length squared, so we take its entries to be
\begin{equation}
  \mathsf{H}_{xx} = \mathsf{H}_{yy} = h^{2}p^{-1}
  \quad \text{and} \quad   \mathsf{H}_{zz} = h^{2}p^{2} ,
  \label{eq:H_uniaxial}
\end{equation}
where $h$ denotes the mesoscopic length scale governing the elastic response of the polymer network.
Although $h$ is related to the network mesh size and, therefore, to the elastomer stiffness, this dependence is not considered in this section where we focus on general phase behavior.

As the elastomer is close to criticality, we set $\phiz \approx \phic$ in Eq.~\eqref{eq:M_general}, and find the longitudinal modulus of a uniaxially swollen elastomer to be
\begin{align}
  M_{\bm{q}} &= \nu\kbt\phic^{-5/3}\left[\phic^{2/3} + \left(p^{2} - p^{-1}\right)\qz + p^{-1}\right]\NN\\
             &\quad\times\exp\left\{-h^{2}q^{2}\left[\left(p^{2} - p^{-1}\right)\qz + p^{-1}\right]\right\},
             \label{eq:M_q_uniaxial}
\end{align}
where $\qz = q_{z}^{2}/q^{2}$.
The presence of the terms proportional to $\qz$ breaks the rotational symmetry of the free-energy functional.%
\footnote{Similar terms appear in the free energy of phase-separating systems with broken symmetries, e.g.,
thin-film diblock copolymers in the presence of a unidirectional electric field~\cite{tsori2002}.}

One can show using elementary calculus that  $\qz = q_{z}^{2}/q^{2}$ is discontinuous at $\bm{q} = 0$ with $\lim_{\bm{q} \to 0} \qz$ failing to exist.
Indeed, if we move along the $q_{z}$-axis towards the origin with $q_{z} \neq 0$ and $q_{x} = q_{y} = 0$, we have $\qz\, {= }\,1$.
On the other hand, on moving towards the origin while remaining on the $(q_{x}, q_{y})$ plane, we have $\qz = 0$.
Therefore, when $p\neq 1$, we see that $M_{\bm{q}}$ also becomes discontinuous at $\bm{q} = 0$, and approaches two different limiting values, with
\begin{alignat}{2}
  \label{eq:M_q1}
  \lim_{q_{z} \to 0} M_{(0, 0, q_{z})} &= M_{\perp} &&= M_{0}\left(\frac{p^{2} + \phic^{2/3}}{1 + \phic^{2/3}}\right), \\
  \label{eq:M_q2}
  \lim_{(q_{x}, q_{y}) \to (0, 0)} M_{(q_{x}, q_{y}, 0)} &= M_{\|} &&= M_{0}\left(\frac{p^{-1} + \phic^{2/3}}{1 + \phic^{2/3}}\right),
\end{alignat}
where $M_{0} = \nu\kbt(\phic^{-1} + \phic^{-5/3})$ is the long-wavelength longitudinal modulus for isotropic swelling,
discussed in detail in Part I~\cite{part1}.
As illustrated in Eqs.~\eqref{eq:M_q1} and \eqref{eq:M_q2}, throughout this paper, we use the subscript $\|$ whenever $q_{z} = 0$ to signify that these modes have wavevectors confined to the $(q_{x},q_{y})$ Fourier plane.
The corresponding spatial modulations are \emph{parallel} to the spatial $z$-axis.
Conversely, the subscript $\perp$ refers to the case $q_{z} \neq 0$ ($q_{x} = q_{y} = 0$), which corresponds to spatial modulations that are \emph{perpendicular} to the spatial $z$-axis.

\subsection{Linear stability analysis}

For linear stability analysis around the uniform state with $\psi = \psiz$, we set $\psi(\bm{x}) = \psiz + \delta\psi(\bm{x})$ in the total free energy and examine the Gaussian free energy $\mathscr{F}_{\text{G}}$, which is quadratic in the modulations $\delta\psi$.
Similar to the isotropic case, it is given by
\begin{align}
  \mathscr{F}_{\text{G}}[\delta\psi] & = \frac{1}{2}\int\frac{\dd^{3}q}{(2\pi)^{3}}\, F_{\bm{q}}\,\delta\psi_{\bm{q}}\delta\psi_{-\bm{q}}, \\
  F_{\bm{q}}                         & = a(T-\Tc) + 3b\psiz^{2} + \kappa q^{2} + M_{\bm{q}}.
  \label{eq:free_binary}
\end{align}
Here, $F_{\bm{q}}$ denotes the Fourier transform of the effective binary interaction governing the modulations $\delta\psi$.
The term $\kappa q^{2}$ in $F_{\bm{q}}$ is a measure of interfacial energy costs and it favors small-$q$ (long-wavelength) modulations.
In contrast, the exponentially decaying elastic term $M_{\bm{q}}$, Eq.~\eqref{eq:M_q_uniaxial}, favors large-$q$ (short-wavelength) modulations.
The competition between these two contributions can therefore stabilize a spatially modulated microphase at an intermediate length scale, provided that the elastic effects dominate.
The characteristic length scale and the spatial structure of the resulting modulation is governed by the most unstable wavevector $\bm{q}_{\text{m}}$.
It can be found, as usual, by minimizing $F_{\bm{q}}$ above~\cite{gennes1979,andelman2009}.
As we show below, the existence of a nonzero $\bqm$ is controlled by an elastocapillary number, which in the isotropic case is given by
\begin{equation}
  \gammaz = \frac{M_{0}h^{2}}{\kappa}.
\end{equation}
The elastocapillary number quantifies the relative importance of elastic and interfacial effects~\cite{ronceray2022}.
Elastic effects dominate over interfacial ones when $\gammaz > 1$ (and vice versa), in which case an isotropically swollen elastomer can undergo microphase separation, as we discussed in Part~I~\cite{part1}.

For uniaxial swelling, owing to the discontinuity of $M_{\bm{q}}$ at $\bm{q} = 0$, direct analysis reveals that the most unstable wavevector $\bm{q}_{\text{m}}$ is either along the $q_{z}$-axis or is confined to the $(q_{x}, q_{y})$ plane.
We discuss those two cases separately.

\begin{figure*}
  \centering\includegraphics{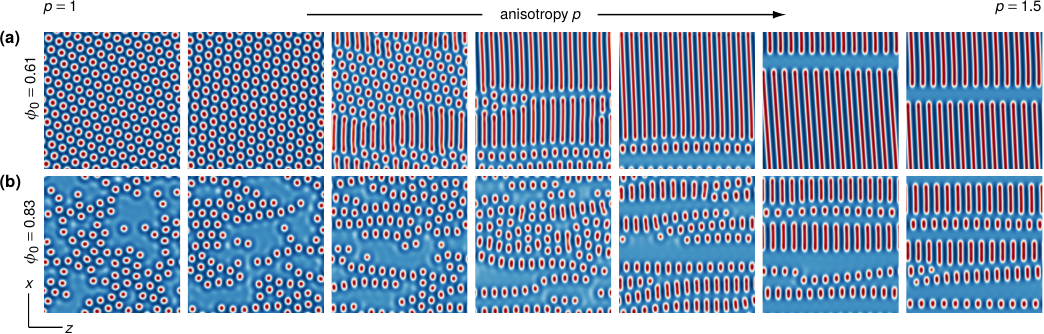}
  \caption{Evolution of the microphase morphology as the anisotropy in the initial swelling increases
    in equal increments from $p = 1$ (isotropic state) to $p = 1.5$ (left to right).
    Polymer-rich and solvent-rich regions are shown in blue and red, respectively.
    In all panels, the temperature $T = -3.4$ and the elastocapillary number $\gammaz = 4$.
    Other parameters are the same as in Fig.~\ref{fig:uniphase}.
    The mean polymer volume fraction $\phiz$ is chosen to be $\phiz = 0.61$ in (a), and $\phiz = 0.83$ in (b).
    For isotropic swelling, they correspond to a hexagonal phase and the hexagonal-uniform phase coexistence (see Fig.~2 of Part I~\cite{part1}).
    As the anisotropy increases, the system undergoes a series of first-order transitions, approximately exhibiting lamellar-hexagonal, lamellar-hexagonal-uniform, and hexagonal-uniform phase coexistence.
    The domain size increases only weakly with $p$.
  }
  \label{fig:weakp}
\end{figure*}

\paragraph{$\bm{q}_{\text{m}}$ along the $q_{z}$-axis.} Substituting Eq.~\eqref{eq:M_q_uniaxial}
in Eq.~\eqref{eq:free_binary}, we find that the unstable wavevector
on the $q_{z}$-axis is $\bm{q}_{\text{m}} = (0,\,0,\,q_{\text{m},z})$, where
\begin{equation}
  q_{\text{m},z}^{2} = h_{\perp}^{-2} \ln \gamma_{\perp}.
  \label{eq:qm_perp}
\end{equation}
The length scale $h_{\perp}$ and the elastocapillary number $\gamma_{\perp}$, defined analogously to the isotropic case, are
\begin{align}
  h_{\perp}      & = ph,                                        \\
  \gamma_{\perp} & = \frac{M_{\perp} h_{\perp}^{2}}{\kappa^{2}}
  = \gammaz p^{2} \left( \frac{p^{2} + \phic^{2/3}}{1 + \phic^{2/3}} \right).
\end{align}
As the wavevector $\bm{q}_{\text{m}}$ lies along the $q_{z}$-axis, the modulations associated with Eq.~\eqref{eq:qm_perp} is a lamellar phase oriented perpendicular to the spatial $z$-axis.
This is true for both two-dimensional (2D) elastomers restricted to the $(z, x)$ plane and 3D elastomers occupying the full $(x, y, z)$ volume.

\paragraph{$\bm{q}_{\text{m}}$ in the $(q_{x}, q_{y})$ plane.} In the $(q_{x}, q_{y})$ plane,
the most unstable wavevector is $\bm{q}_{\text{m}} = (q_{\text{m},x},\, q_{\text{m},y},\, 0)$,
with its components lying on a circle given by
\begin{equation}
  q_{\text{m},x}^{2} + q_{\text{m},y}^{2} = h_{\|}^{-2} \ln \gamma_{\|} ,
  \label{eq:qm_par}
\end{equation}
and $h_{\|}$ and $\gamma_{\|}$ are given by
\begin{align}
  h_{\|}      & = p^{-1/2} h,\\
  \gamma_{\|} & = \frac{M_{\|} h_{\|}^{2}}{\kappa^{2}}
  = \gammaz p^{-1} \left( \frac{p^{-1} + \phic^{2/3}}{1 + \phic^{2/3}} \right).
\end{align}
In 3D, the modulations associated with Eq.~\eqref{eq:qm_par}, are either a parallel lamellar phase with $q_{x} \neq 0, q_{y} = 0$ (or $q_{x} = 0, q_{y} \neq 0$) or a hexagonally packed cylindrical phase ($q_{x}, q_{y} \neq 0$) oriented parallel to the spatial $z$-axis.
However, in 2D, only lamellae emerge.

Similar to the isotropic case, the wavevectors described by Eqs.~\eqref{eq:qm_perp} and \eqref{eq:qm_par} only exist if the corresponding elastocapillary numbers $\gamma_{\perp}$ and $\gamma_{\|}$ are larger than unity.
If such wavevectors exist, they become unstable when the temperature falls below the temperature at which $F_{\bm{q}}$ first vanishes.
This defines the microphase separation temperature,
\begin{equation}
  \label{eq:Tm}
  \Tm(\psiz) =
  \Tc - a^{-1}\left[3b\psiz^{2} + M_{\alpha}\gamma_{\alpha}^{-1}(1 + \ln\gamma_{\alpha})\right],
\end{equation}
where the index $\alpha$ is $\perp$ or $\|$, denoting the two orientations.

When $\gamma_{\perp}<1$ or $\gamma_{\|}<1$, the corresponding finite-wavevector instability does not exist.
Consequently, the system does not undergo microphase separation.
Instead, the first unstable mode is the homogeneous mode, $\bm{q}=0$, which becomes unstable below the temperature
\begin{equation}
  T_{0}(\psiz) = \Tc - a^{-1}(3b\psiz^{2} + M_{\alpha}), \quad \alpha =\, \perp,\, \|.
  \label{eq:Tmacro}
\end{equation}
Unlike the microphase separation temperature $\Tm$, at which spatially modulated structures emerge, $T_{0}$ marks the onset of a bulk instability.
Below $T_{0}$, the system undergoes macrophase demixing between the solvent and the polymer network rather than forming a periodically modulated microphase.

\begin{figure*}
  \centering\includegraphics{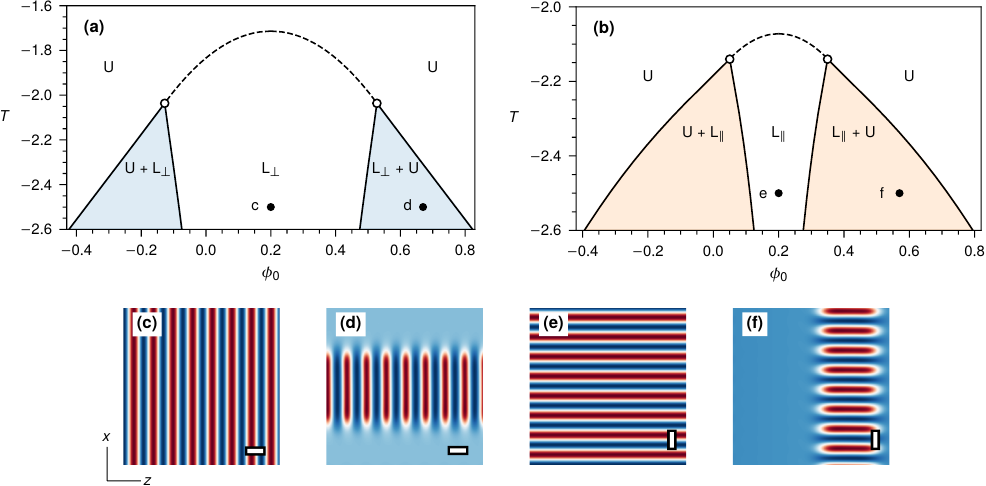}
  \caption{Phase diagram in the polymer volume fraction--temperature $(\phiz, T)$ plane for a uniaxially deformed elastomer
    with elastocapillary number $\gammaz = 4$ and anisotropy factor (a) $p = 3/2>1$ (elongation) and (b) $p = 2/3<1$ (compression).
    For illustrative purposes, we took the critical parameters $(\phic, \Tc) = (0.2, 0)$, and set all other parameters in the free energy to unity.
    Other parameters are the same as in Fig.~\ref{fig:uniphase}.
    In both panels (a) and (b), the dashed curves indicate a line of second-order transitions from the uniform phase to the perpendicular (L$_\perp$) or parallel (L$_\|$) lamellar phase, respectively.
    The colored regions bounded by first-order transition lines (solid curves) show coexistence between the lamellar and uniform phases.
    These curves meet at two tricritical points (open circles).
    (c)--(f) Representative equilibrium morphologies for elongation and compression.
    The $(\phiz, T)$ values used are marked in panels (a) and (b) by a black dot.
    Polymer-rich regions ($\phi>\phiz$) are shown in blue, while solvent-rich regions ($\phi < \phiz$) are shown in red.
    The scale bar indicates the domain size estimated from the magnitudes of the wavevectors in Eqs.~\eqref{eq:qm_perp} and \eqref{eq:qm_par}.
  }
  \label{fig:lamella}
\end{figure*}

\subsection{General phase behavior}

By numerically comparing the values of $F_{\bm{q}}$ below the transition temperatures given in Eqs.~\eqref{eq:Tm} and \eqref{eq:Tmacro}, we can construct the phase diagram describing the morphologies that first emerge during a temperature quench at the critical volume fraction $\phic$ (or, equivalently, for order parameter $\psiz = 0)$.
Such a phase diagram should be regarded as approximate, since it is constructed solely based on the system's linear stability.
Independent of the specific values of the physical parameters appearing in the free energy, this phase diagram can be represented entirely in the $(p, \gammaz)$ plane.

An example of such a phase diagram is shown in Fig.~\ref{fig:uniphase}(a), and it has four distinct regions corresponding to microphase separation.
For $\gammaz \gtrsim 1$, uniaxial extension ($p > 1$) produces a lamellar phase aligned parallel to the $z$-axis (L$_\parallel$), whereas uniaxial compression ($p < 1$) leads to a lamellar phase aligned perpendicular to the $z$-axis (L$_\perp$).
In the limit $\gammaz \gg 1$, this behavior is reversed: uniaxial compression yields the parallel lamellar phase L$_{\|}$, while uniaxial extension results in the perpendicular lamellar phase L$_\perp$.

The four phases within the microphase-forming region are separated by curves analogous to second-order transition lines, across which the average value of the order parameter remains fixed at $\psiz = 0$.
Below $\gammaz = 1$, microphase separation is not possible, and the system undergoes macrophase separation into solvent-rich and solvent-poor phases.
Thus, similar to an isotropically swollen elastomer~\cite{part1}, the parameter value $\gammaz = 1$ corresponds to the Lifshitz point~\cite{hornreich1975} of the uniaxially deformed elastomer.

In the two-phase coexistence region, domain growth during early stages of macrophase separation, would occur either parallel or perpendicular to the $z$-axis.
This depends on the relative values of $M_{\perp}$ and $M_{\|}$, Eqs.~\eqref{eq:M_q1} and \eqref{eq:M_q2}.
For $p > 1$, we have $M_{\|} < M_{\perp}$, and one would observe lamellar or cylindrical domains that grow parallel to the spatial $z$-axis.
When $p < 1$, the situation is reversed, and the growth of the lamellar domains is perpendicular to the $z$-axis.
The behavior of uniaxially deformed gels~\cite{onuki1993} and metallic alloys~\cite{onuki1989a} undergoing spinodal decomposition is rather similar.
The microphase-forming regions are separated from the two-phase regions by curves analogous to triple lines.

The predictions of the phase diagram in Fig.~\ref{fig:uniphase}(a) are in good agreement with the morphologies of the phase-separated domains showcased
in Fig.~\ref{fig:uniphase}(b)--\ref{fig:uniphase}(g).
These morphologies were obtained by numerically minimizing the total free energy in a 2D box using Model B dynamics~\cite{bray1994},
as described in Appendix~\ref{app:numerical}.

Let us now turn our attention to off-critical ($\psiz= \phiz-\phic \neq 0$) phase behavior.
The phase-separated elastomer as the anisotropy increases from $p = 1$ (isotropic state) to $p = 1.5$ is shown in Fig.~\ref{fig:weakp} for two different values of the mean polymer volume fraction, $\phiz$.
In Fig.~\ref{fig:weakp}(a), the equilibrium morphology for an initial isotropic state is a hexagonal phase.
However, in Fig.~\ref{fig:weakp}(b), for an initial isotropic state, the system is in a coexistence region between the hexagonal and uniform phases.
As can be seen in Fig.~\ref{fig:weakp}(a), when $p\gtrsim 1$, the system continues to form hexagonally symmetric domains, similar to the isotropic case (discussed in Part~I~\cite{part1}).
As $p$ increases further, the hexagonal phase gradually evolves into the perpendicular%
\footnote{Parallel lamellae do not appear as the elastocapillary number $\gammaz = 4$ in these figures.}
lamellar phase L$_\perp$, with a narrow range of $p$ values where the L$_{\perp}$ and hexagonal phases coexist.

At even larger $p$, as we see from Fig.~\ref{fig:weakp}(a), the hexagonal phase completely disappears, and one sees a coexistence between the L$_{\perp}$ and uniform phases.
This trend is generally observed even when starting in the uniform-hexagonal phase coexistence region and slowly increasing $p$ as in Fig.~\ref{fig:weakp}(b).
However, in this situation, at intermediate $p$ values, one can also observe coexistence among the uniform, hexagonal, and lamellar phases, with the hexagonal phase disappearing as $p$ increases.
Finally, for sufficiently strong anisotropy, irrespective of the value of $\phiz$ the system prefers to be in a lamellar phase or to remain in the uniform phase.

\subsection{Strong anisotropy}

Phase diagrams can be obtained analytically for strong anisotropy ($p \gg 1$ and $p \ll 1$) by using the single-mode approximation.
For simplicity, we restrict our analysis to 2D elastomers and assume the existence of only two phases---the lamellar phase (either L$_{\perp}$ or L$_{\|}$) and the uniform phase.
Hexagonal phases are excluded as they only appear in 3D (only when $q_{x}, q_{y} \neq 0$ and $q_{z} = 0$).
See Part~I~\cite{part1} for a general overview of the theory of modulated phases.

The free-energy density $f_{\text{L}}$ for the lamellar phase can be obtained by putting $\psi(\bm{x}) = \psiz + A\cos(\bm{q}_{\text{m}}\cdot\bm{x})$ in the total free energy, Eq.~\eqref{eq:free_total_simple}, and minimizing it with respect to the amplitude $A$.
This leads to
\begin{align}
  f_{\text{L}}(\psiz, T) & = f_{\text{U}}(\psiz, T) - (6b)^{-1}\big[a(T-\Tc)\NN                                  \\
                         & \quad+ 3b\psiz^{2} + M_{\alpha}\gamma_{\alpha}^{-1}(1 + \ln\gamma_{\alpha})\big]^{2},
  \label{eq:free_lam_anis}
\end{align}
with the free-energy density of the uniform phase being
\begin{equation}
  f_{\text{U}}(\psiz, T) = \frac{1}{2}\left[a(T-\Tc) + M_{\beta}\right]\psiz^{2} + \frac{1}{4}b\psiz^{4}.
  \label{eq:free_unif_anis}
\end{equation}
In these expressions, the subscripts $\alpha, \beta$ are $\perp$ or $\|$ as in Eqs.~\eqref{eq:M_q1} and \eqref{eq:M_q2} for the longitudinal modulus.
We use different subscripts in $f_{\text{U}}$ and $f_{\text{L}}$ to emphasize that, depending on where we are on the $(p,\gammaz)$ phase plane, the longitudinal modulus $M_\beta$ of the uniform phase in Eq.~\eqref{eq:free_unif_anis} may differ from that of the lamellar modulations in Eq.~\eqref{eq:free_lam_anis}.
Indeed, in the limit $q \to 0$, the value of $M_{\bm{q}}$ can be either $M_{\|}$ or $M_{\perp}$, and the system always selects the one that minimizes the total free energy.
In particular, when $p>1$, we have $M_{\perp}>M_{\|}$.
Therefore, regardless of whether the lamellar modulations are oriented parallel or perpendicular to the $z$-axis, when $p > 1$, the longitudinal modulus of the uniform phase is $M_{\|}$---even when there is phase coexistence with the L$_\perp$ phase.
Note that the situation is reversed for $p < 1$.

The phase diagram obtained from a common-tangent construction of the free-energy densities in Eqs.~\eqref{eq:free_lam_anis} and \eqref{eq:free_unif_anis} is shown in
Fig.~\ref{fig:lamella}(a) for $p > 1$ and in Fig.~\ref{fig:lamella}(b) for $p < 1$.
Just below the transition temperature, the system can undergo a second-order transition
(no discontinuous change in the order parameter) from a uniform phase to the L$_\perp$ phase (for $p > 1$) or to the parallel one, L$_\parallel$ (for $p < 1$).
The phase transition becomes first order below a tricritical point, denoted by an empty circle in Fig.~\ref{fig:lamella}.
Below the tricritical point, the uniform and lamellar phases coexist.
The morphologies predicted by the phase diagrams agree rather well with those obtained via numerical minimization
of the free energy and are depicted in Fig.~\ref{fig:lamella}(c)--\ref{fig:lamella}(f).
It should be emphasized that in 3D, Fig.~\ref{fig:lamella}(b) would capture the phase behavior only near the critical volume fraction.
Away from it, one would expect the emergence of hexagonally symmetric cylindrical domains oriented along the spatial $z$-axis, which we do not consider in our analysis.

A line of second-order phase transitions, like the dashed ones in Fig.~\ref{fig:lamella}, also exists between the uniform and lamellar phases in the case of isotropic swelling.
Nevertheless, when considering 2D or 3D elastomers, such a line always lies within the boundaries of other phases and does not exist in thermodynamic equilibrium.
Consequently, a second-order phase transition to patterned phases is not observed for isotropically swollen elastomers in 2D or 3D.
However, an isotropically swollen elastomer in 1D can exhibit a second-order phase transition from a uniform state to a patterned, lamellar phase~\cite{mannattil2025}.
This behavior is also seen in the 1D version of the phase-field crystal model~\cite{elder2004,thiele2019}, and the Blume--Emery--Griffiths model originally applied to $^{3}$He-$^{4}$He mixtures~\cite{chaikin1995} and later used to model other phenomena, e.g., membrane adhesion~\cite{komura2000}.

\begin{figure*}
  \centering\includegraphics{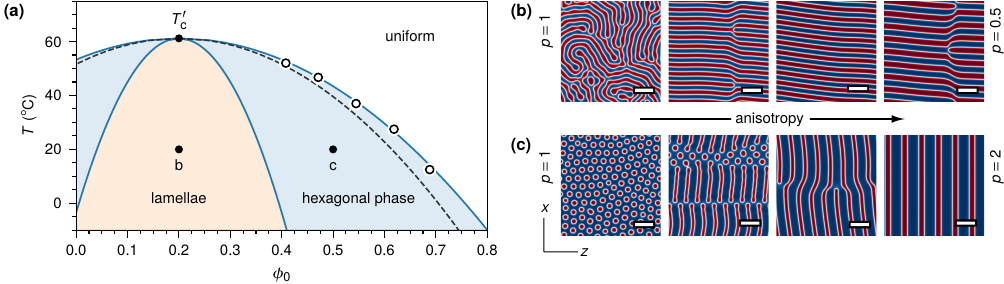}
  \caption{Phase diagram in the $(\phiz,T)$ plane for an isotropically swollen elastomer with dry-state Young's modulus $Y=350\unit{\text{kPa}}$.
    The dashed curve marks the microphase-separation boundary predicted by linear stability analysis.
    The binodals (solid curves) terminate at a shifted critical point $\Tc'$.
    For constructing the phase diagram, we chose the parameter values $\kappa = 0.013\unit\text{kPa}\unit\text{\textmu m}^{2}$, $\Tc = 70\unit^{\circ}\text{C}$, $\phic=0.2$, $a=0.025\unit\text{kPa}\unit\text{K}^{-1}$, and $b = 2\unit\text{kPa}$ in the free energy~\cite{part1}.
    Representative morphologies obtained via numerical minimization of the free energy are shown in panels (b) and (c).
    The associated $(\phiz, T)$ values are marked by black dots in panel~(a).
    In both (b) and (c), the degree of anisotropy increases from left to right due to uniaxial compression or extension along the $z$-axis.
    As a result, the morphologies evolve toward lamellae aligned parallel (L$_{\|}$) or perpendicular (L$_{\perp}$) to the $z$-axis.
    Polymer-rich regions $(\phi > \phiz)$ are shown in blue, while solvent-rich regions $(\phi < \phiz)$ are shown in red.
    The scale bar is $4\unit\text{{\textmu}m}$.
  }
  \label{fig:comparison}
\end{figure*}

\section{Comparison with experiments}
\label{sec:experiments}

Following Part~I~\cite{part1}, we assume that the characteristic length scale governing continuum elasticity is proportional to the end-to-end distance $\xi$ of strands between neighboring crosslinks of the polymer network.
The distance $\xi$ is a measure of the network's mesh size~\cite{canal1989,yoo2006,parrish2017,richbourg2020}.
Modeling each strand as a freely jointed chain with Flory ratio $C_{\infty}$, consisting of $N$ monomers of length $\ell$, gives $\xi^{2}\sim C_{\infty}N\ell^{2}$~\cite{tanaka2011,richbourg2020}.
If $\ms$ and $\mz$ denote the molecular masses of a strand and monomer, respectively, then $N=\ms/\mz$.
For an elastomer of mass density $\varrho$, the dry-state Young's modulus is given by $Y=3\nu\kbt$ where the strand density $\nu = \varrho/\ms$~\cite{lodge2020}.
Eliminating $\mz$ in favor of $Y$ therefore yields the scaling relation for the mesh size $\xi \sim (3B/Y)^{1/2}$.
Here $B=C_{\infty}\varrho\ell^{2}\kbt/\mz$ is a material parameter.
For physical parameters relevant to the PDMS elastomers used in Ref.~\cite{fernandez-rico2024}, we estimate $B=0.024\unit\text{kPa}\unit\text{\textmu m}^{2}$.
See Part~I~\cite{part1} for more detailed discussions on these parameter values.

Continuum elasticity becomes applicable only at length scales substantially larger than the mesh size $\xi$.
We therefore choose the characteristic length scale $h$ to be proportional to the mesh size: $h = n\xi\phic^{-1/3}$.
Here, $n$ represents the effective number of crosslinks we coarse-grain over.
It is treated as a fitting parameter and we choose $n=35$ as in the isotropic case.
The factor of $\phic^{-1/3}$ in $\xi$ accounts for swelling.
The longitudinal modulus $M_{\bm{q}}$ in Eq.~\eqref{eq:M_general} is expressed in terms of the dry Young's modulus $Y$ after eliminating the strand density $\nu$ from $Y = 3\nu\kbt$~\cite{lodge2020}.
For isotropically swollen elastomers, writing $\xi$ and $M$ in terms of $Y$ leads to the size of the phase-separated domains scaling as $Y^{-1/2}$~\cite{part1}.

Unlike isotropically swollen elastomers, the available results for EMPS in anisotropic elastomers are rather limited and qualitative.
Experiments of Ref.~\cite{fernandez-rico2024} have mainly focused on two types of anisotropy: the influence of a stiffness gradient on EMPS and the effects of uniaxial swelling.

\subsection{Uniaxial swelling}

For uniaxial swelling along the spatial $z$-axis, the length covariance matrix $\mathsf{H}$ in Eq.~\eqref{eq:H_uniaxial} is diagonal with entries $h^{2}p^{-1}$, $h^{2}p^{-1}$, and $h^{2}p^{2}$ with $p$ quantifying the degree of anisotropy.
Figure~\ref{fig:comparison} illustrates the effect that uniaxial swelling has on the morphologies of the phase-separated domains for an elastomer of stiffness $Y = 350\unit\text{kPa}$.
The phase diagram for isotropic swelling is included for reference and includes both lamellar and hexagonal phases~\cite{part1}.
Under uniaxial compression, the system undergoes a transition towards lamellar structures that are oriented parallel to the axis of deformation (L$_\parallel$), as shown in Fig.~\ref{fig:comparison}(b).
In comparison, under uniaxial extension, the lamellae are oriented perpendicular to this axis (L$_\perp$) as in Fig.~\ref{fig:comparison}(c).
This behavior is consistent with theoretical expectations, as the elastocapillary number $\gammaz$ is large for these elastomers.

The qualitative description of the two representative morphologies reported in Ref.~\cite{fernandez-rico2024} suggests a trend that differs from our theoretical prediction.
In the experiments, under uniaxial compression, the phase-separated domains appear compressed along the direction of the compression.
Likewise, when the elastomer is stretched, the domains get stretched in the same direction.
This suggests that other mechanisms are likely involved.
For instance, if the initial swelling is strongly inhomogeneous, then it cannot be considered to be affine~\cite{onuki2002}.
A rather similar discrepancy also arises in the so-called ``butterfly effect'' seen in crosslinked polymer blends and gels that are uniaxially deformed~\cite{rouf1994}.
In these systems, even though there is no microphase separation, the scattering intensity is sometimes seen to increase along the stretched direction, whereas theoretical models that assume homogeneous swelling predict the opposite~\cite{onuki1992,panyukov1996}.

\begin{figure*}
  \centering\includegraphics{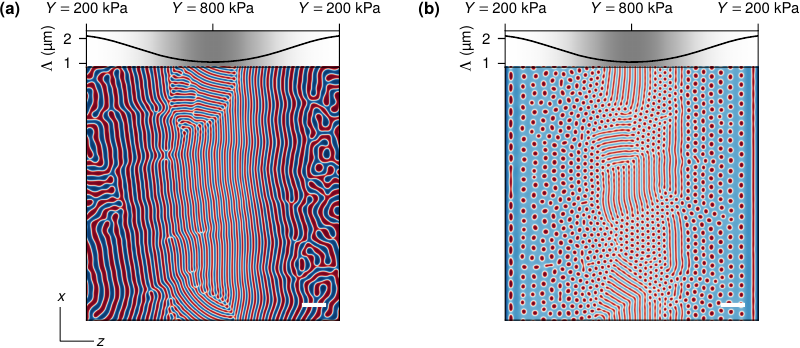}
  \caption{Morphologies of the microphase-separated domains in an elastomer with varying stiffness
  when the mean polymer volume fraction is (a) at criticality, $\phiz =\phic= 0.2$ and (b) off critical, $\phiz = 0.5>\phic$.
   Polymer-rich and solvent-rich regions are colored in blue and red, respectively, and the scale bar is $4\unit\text{\textmu m}$.
  In both (a) and (b), the dry Young's modulus varies from $Y = 200\unit\text{kPa}$ to $Y = 800\unit\text{kPa}$ (as indicated by the gradient in the top panels).
  This leads to a predicted domain size
  $\Lambda = 2\pi\qm^{-1} \approx 1\text{--}2\unit\text{\textmu m}$ (solid curves in the top panels),
  in agreement with the experimental observations in Ref.~\cite{fernandez-rico2024}.}
  \label{fig:varstiff}
\end{figure*}

\subsection{Stiffness gradients}

Let us consider the effect of a stiffness gradient on the elastomer network.
In experiments, such an elastomer can be fabricated, for example, by sandwiching partially cured polymer-crosslinker mixtures of different stiffnesses~\cite{fernandez-rico2024}.
We do not analytically study the detailed phase behavior of these systems because the resulting morphologies cannot be represented as simple sinusoidal modulations.
Although the morphologies are generally complex, we expect lamellar patterns to emerge at least near the critical volume fraction $\phic$.
Away from $\phic$, in 2D, hexagonal-like modulations are expected to appear, at least locally.
For off-critical quenches in 2D, the final morphology is, therefore, anticipated to consist of a mixture of lamellar and hexagonal-like structures, with a local domain size inversely varying with the local stiffness.

The morphologies obtained via numerical minimization of the free energy after introducing stiffness gradients are shown in Fig.~\ref{fig:varstiff} (see Appendix~\ref{app:numerical} for details on the numerical procedures).
Here we have not assumed anisotropy during swelling, which is taken to be \emph{locally} affine and isotropic.
A stiffness gradient was imposed by choosing the local crosslink density to be a smoothly varying function of the $z$ coordinate.
This makes the long-wavelength longitudinal modulus $M_{0}(\bm{x})$, Eq.~\eqref{eq:M_general}, as well as the length scale $h(\bm{x})$ position dependent.

Figure~\ref{fig:varstiff} also shows the local domain size predicted by linear stability analysis, which scales with the Young's modulus as $Y^{-1/2}$~\cite{part1}.
As anticipated, softer regions exhibit larger domains, while stiffer regions develop smaller domains.
This spatial variation in domain size is also consistent with the experimental observations reported in Ref.~\cite{fernandez-rico2024} for elastomers with a stiffness gradient.

\section{Concluding remarks}
\label{sec:conclusion}

In this study (Part~II of a two-part paper), we have extended the theory developed in Part~I~\cite{part1} to account for anisotropic effects.
We employed linear elasticity obtained by linearizing nonlinear rubber elasticity about anisotropically swollen states, along with the nonlocal approach developed in Part~I.
Anisotropic effects are difficult to incorporate within phenomenological descriptions of elastic microphase separation (EMPS) based on alternative theories of elasticity, such as the one used in our earlier work~\cite{mannattil2025}.

EMPS in isotropically swollen elastomers is expected to be a first-order phase transition.
However, we find that uniaxially swollen elastomers can instead undergo a continuous transition between the uniform and lamellar phases.
This result is surprising because continuous lamellar--uniform transitions are generally confined to 1D systems and are not expected in their higher-dimensional counterparts~\cite{thiele2019,elder2004}.

In the case of anisotropically swollen elastomers, for physical parameters relevant to the experiments, we expect lamellar morphologies after phase separation with the lamellae oriented parallel to the axis along which the elastomer is uniaxially compressed, and perpendicular to the axis when it is stretched (see Fig.~\ref{fig:comparison}).
Unfortunately, only limited experimental results are available for such anisotropically deformed systems.
The two sample morphologies reported in Ref.~\cite{fernandez-rico2024} deviate from the theoretical predictions: uniaxial compression along a particular direction results in patterns compressed along that direction, and vice versa.
Current experimental data on EMPS do not provide sufficient guidance to explain this contradiction, suggesting the presence of additional physics beyond what we have considered here.

Our predictions for elastomers with a stiffness gradient are consistent with the available experimental observations, showing that smaller microphase domains form in stiffer regions, while larger domains develop in softer regions.
This demonstrates that spatially varying stiffness is an effective strategy for controlling the local microstructure and highlights EMPS as a versatile route to producing functionally graded materials~\cite{miyamoto1999}.
A natural extension of the present framework would be to incorporate nonaffine, spatially inhomogeneous swelling ~\cite{onuki1992} and to consider the effect of network heterogeneities and quenched impurities~\cite{zhao2009,ghosh2020,ghosh2024}.
Such effects may significantly influence the resulting phase behavior and morphology.
Exploring these effects remains an interesting direction for future work.

\subsection*{Acknowledgments}

Conversations with L.~Mahadevan, Mehrana Nejad, and Sam Safran have been fruitful.
H.D.~acknowledges support from the Israel Science Foundation (ISF Grant No.~1611/24).
D.A.~acknowledges support from the Israel Science Foundation (ISF Grant No.~226/24).

\appendix

\section{Continuum mechanics of elastomers}
\label{app:continuum}

\subsection{Kinematics}

Using standard notations from continuum mechanics, we represent the dry, unswollen elastomer in 3D by the material (Lagrangian) coordinates $\bm{X}$.
We assume that the elastomer deforms affinely when it swells and take the spatial (Eulerian) coordinates $\bm{x}_{0}$ of the swollen elastomer to be
\begin{equation}
  \bm{x}_{0} = \mathsf{A}\bm{X},
\end{equation}
where $\mathsf{A}$ is an invertible $3\times 3$ matrix with constant entries representing the affine deformation associated with swelling.
If phase separation occurs, the elastomer deforms again, and the final coordinates $\bm{x}$
describing the equilibrium state of the elastomer after phase separation are
\begin{equation}
  \bm{x}(\bm{X}) = \bm{x}_{0}(\bm{X}) + \bm{u}(\bm{X}) = \mathsf{A}\bm{X} + \bm{u}(\bm{X}),
  \label{app:eq:xX}
\end{equation}
where the displacement field $\bm{u}(\bm{X})$, measured relative to the affinely swollen state,
captures the additional deformation that occurs during phase separation.
Throughout, the coordinates $\bm{X}$, $\bm{x}_{0}$, and $\bm{x}$ describe only the polymer network in the elastomer, not the solvent.

The volume elements in the spatial coordinates, $\dd^{3}x_{0}$ and $\dd^{3}x$, are related to the volume element $\dd^{3}X$ in the material coordinates via
\begin{equation}
  \dd^{3}x_{0} = J_{0}\,\dd^{3}X
  \quad
  \text{and}
  \quad
  \dd^{3}x = J(\bm{X})\,\dd^{3}X ,
  \label{app:eq:vol_elements}
\end{equation}
where $J_{0}$ and $J(\bm{X})$ are the Jacobian determinants of the coordinate transformations $\bm{X} \to \bm{x}_{0}$ and $\bm{X} \to \bm{x}$, respectively.
As $\mathsf{A}$ is a constant matrix, $J_{0} = \det\mathsf{A}$ is also constant.
On the other hand, $J = \det\mathsf{F}$ where
\begin{equation}
  \mathsf{F} = \mathsf{A} + \gradX\bm{u}(\bm{X})
\end{equation}
is the deformation gradient tensor.
In what follows, we assume the displacement field $\bm{u}$ to be small.
So, to leading order in $\bm{u}$, we have
\begin{align}
  J = \det\mathsf{F} &\approx \det\left[\mathsf{A} + (\grad\bm{u})\mathsf{A}\right]\NN\\
                     &= \det\left(\mathbbm{1} + \grad\bm{u}\right)\det\mathsf{A}\NN\\
                     &\approx J_{0}\left[1 + \div\bm{u} + \mathcal{O}(\abs{\bm{u}}^{2})\right],
                     \label{app:eq:jacobian}
\end{align}
where, in the first line, we have made use of $\partial_{X_{j}} = (\partial x_{k}/\partial X_{j}) \partial_{x_{k}} =
\mathsf{A}_{kj} \partial_{k} + \mathcal{O}(\bm{u})$ to express all partial derivatives of $\bm{u}$ in terms of the spatial coordinates $\bm{x}$.
As usual, repeated indices are summed over.

Let $\phiz$ and $\phi(\bm{x})$ denote the polymer volume fractions immediately after swelling and after the phase separation, respectively.
The total polymer volume is conserved during both swelling and phase separation.
Assuming that the polymer volume fraction is unity in the dry state, this implies a \emph{local}
material conservation relation of the form~\cite{mannattil2025,reddy2013}
\begin{equation}
  J_{0}\phiz = J(\bm{x})\bar{\phi}(\bm{x}) = 1,
  \label{app:eq:matcons}
\end{equation}
where $\bar{\phi}(\bm{x})$ is the coarse-grained polymer volume fraction obtained from $\phi(\bm{x})$ via Eq.~\eqref{eq:blur}.
Here, we use $\bar{\phi}(\bm{x})$ rather than $\phi(\bm{x})$ to ensure that the elastic deformations of the polymer network
are associated only with variations in $\phi(\bm{x})$, which occur above a mesoscopic length scale.
Also, as $\phiz$ is a constant, $\bar{\phi}_{0} = \phiz$.

Assuming small deformations close to the critical point, we can expand in Eq.~\eqref{app:eq:matcons} both
$\bar{\phi}(\bm{x})$ and $\phiz$ around $\phi = \phic$.
Upon using the resulting expression in Eq.~\eqref{app:eq:jacobian}, we find
\begin{equation}
  \div\bm{u} = -\phic^{-1}\bar{\psi} + \mathcal{O}(\psi_{0}) + \mathcal{O}(\psi^{2}),
\end{equation}
where $\bar{\psi}(\bm{x})=\bar{\phi}(\bm{x}) - \phic$ and $\psi_0 = \phiz - \phic$.

Even though the total volume of the polymer network remains conserved at all stages, the total volume of the system
(which includes the network and solvent) changes during swelling because of solvent uptake.
By contrast, the overall system volume remains unchanged during phase separation since no solvent enters or leaves the system.
Equating the total volume of the system before and after phase separation leads to a \emph{global} volume conservation relationship of the form
\begin{equation}
  \int\dd^{3}x = \int\dd^{3}x_{0} =
  \int\dd^{3}x\,J^{-1}(\bm{x})J_{0},
  \label{app:eq:volcons}
\end{equation}
where we have made use of Eq.~\eqref{app:eq:vol_elements} to relate the volume elements, and the integral is over the space occupied by the system.
Note that, in general, $J(\bm{x}) \neq J_{0}$ as phase separation does not \emph{locally} preserve the volume.
Using Eq.~\eqref{app:eq:matcons}, we can write Eq.~\eqref{app:eq:volcons} as an integral constraint of the form
\begin{equation}
  \int\dd^{3}x\, \phiz^{-1}\left[\phi(\bm{x}) - \phiz\right] = 0.
\end{equation}
The above integral, when expressed in terms of the order parameter $\psi(\bm{x})$, becomes the linear term in the Ginzburg--Landau energy
$\mathscr{F}_{\text{GL}}$, Eq.~\eqref{eq:free_GL}, with the chemical potential $\eta$ serving as a Lagrange multiplier.

\subsection{Energetics}

The strain-energy density $W$ in Eq.~\eqref{eq:energy_density} when written in terms of Jacobian $J$ and the first invariant
$I_{1} = \trace{\mathsf{B}}$ of the left Cauchy--Green deformation tensor $\mathsf{B} = \mathsf{F}\mathsf{F}\trans$ takes the form
\begin{equation}
  W = \frac{1}{2}\nu\kbt(\mathsf{F}_{jk}\mathsf{F}_{jk} - 2\ln{J} - 3).
\end{equation}
As we want to express the elastic energy in the spatial coordinates $\bm{x}$, we consider the Cauchy stress tensor $\stress$,
which quantifies the internal stresses within a hyperelastic material in its deformed state.
Its components $\sigma_{jk}$ are~\cite{reddy2013}
\begin{equation}
  \sigma_{jk} = J^{-1}\frac{\partial W}{\partial F_{jl}} F_{kl}.
\end{equation}

From the Cauchy stress $\stress$ and the linearized strain $\strain = \frac{1}{2}\left[\grad\bm{u} + (\grad\bm{u})\trans\right]$,
we can compute the linearized elastic energy as
\begin{align}
  \mathscr{F}_{\text{el}}[\strain] &= \frac{1}{2}\int\dd^{3}x\,\sigma_{jk}\varepsilon_{jk}\NN\\
                                   &= \frac{1}{2}\nu\kbt \phiz \int\dd^{3}x\,\bigg[(\mathsf{A}_{jl}\mathsf{A}_{kl} - \delta_{jk})\varepsilon_{jk}\\
                                   &\quad- (\mathsf{A}_{jl}\mathsf{A}_{kl} - \delta_{jk})\frac{\partial u_{m}}{\partial x_{m}}\varepsilon_{jk} +
                                   2\mathsf{A}_{jm}\mathsf{A}_{lm}\frac{\partial u_{k}}{\partial x_{l}}\varepsilon_{jk}\bigg]\NN.
\end{align}
Here, we have made use of Eqs.~\eqref{app:eq:jacobian} and \eqref{app:eq:matcons} and have written all derivatives in terms of $\bm{x}$.
The first term in the elastic energy density above is the mechanical part of the osmotic stress from the initial affine swelling.
Owing to the symmetry of the terms involved, we can write it as the total divergence $\partial_{j}[(\mathsf{A}_{jl}\mathsf{A}_{kl} - \delta_{jk})u_{k}]$.
This term becomes a surface term upon integration and can be discarded.
After some straightforward algebra we can use the Parseval--Plancherel identity to write $\mathscr{F}_{\text{el}}$ in Fourier space as
\begin{align}
  \mathscr{F}_{\text{el}}[\bm{u}] &= \frac{1}{2}\nu\kbt \phiz\int \frac{\dd^3{q}}{(2\pi)^{3}}\big[(\bm{q}\cdot\bm{u}_{\bm{q}})(\bm{q}\cdot\bm{u}_{-\bm{q}})\NN\\
                                  &\quad+ \mathsf{A}_{jl}\mathsf{A}_{kl}q_{j}q_{k}(\bm{u}_{\bm{q}}\cdot\bm{u}_{-\bm{q}})\big].
\end{align}
This completes the derivation of Eq.~\eqref{eq:free_el}.

Similar equations for the elastic free energy have been derived by other authors using somewhat different arguments~\cite{onuki1993,panyukov1996}.
Separating $\bm{u}_{\bm{q}}$ into its transverse and longitudinal components, we can write $\mathscr{F}_{\text{el}}$ as
\begin{align}
  \mathscr{F}_{\text{el}}[\bm{u}] &= \frac{1}{2}\nu\kbt \phiz\int \frac{\dd^3{q}}{(2\pi)^{3}}\bigg[
                                     \mathsf{A}_{jl}\mathsf{A}_{kl}q_{j}q_{k}\bm{u}_{\bm{q}}^{\perp}\cdot\bm{u}_{-\bm{q}}^{\perp}\NN\\
                                  &\quad+ \left(1 +
                                  \mathsf{A}_{jl}\mathsf{A}_{kl}\frac{q_{j}q_{k}}{q^{2}}\right)(\bm{q}\cdot\bm{u}_{\bm{q}})(\bm{q}\cdot\bm{u}_{-\bm{q}})\bigg],
\end{align}
where $\bm{u}_{\bm{q}}^{\perp} = \bm{u}_{\bm{q}} - (\hat{\bm{q}}\cdot\bm{u}_{\bm{q}})\hat{\bm{q}}$ is the transverse component of $\bm{u}_{\bm{q}}$,
and $\hat{\bm{q}}=q^{-1}\bm{q}$.
During phase separation, changes in the polymer volume fraction are driven primarily by solvent diffusion,
which is not expected to influence the transverse shear modes $\bm{u}_{\bm{q}}^{\perp}$.
This allows us to discard them, and after using the material conservation relation, Eq.~\eqref{eq:matcons}, we can express the total elastic energy
in terms of the order parameter $\psi$ to obtain the elastic energy in Eq.~\eqref{eq:free_total_simple}.

\section{Numerical techniques}
\label{app:numerical}

In order to numerically minimize the free energy, we make the order parameter $\psi(\bm{x}, t)$ dependent on a fictitious time $t$.
The total free energy $\mathscr{F}$ can then be minimized by evolving $\psi(\bm{x}, t)$ using Model B dynamics~\cite{bray1994}
\begin{equation}
  \frac{\partial \psi(\bm{x}, t)}{\partial t} = \nabla^{2}\eta,\quad
  \eta = \frac{\delta\mathscr{F}}{\delta\psi}.
  \label{app:eq:modelb}
\end{equation}
To evaluate the chemical potential $\eta$, we need to compute the coarse-grained order parameter $\bar{\psi}(\bm{x}, t)$.
If the length covariance matrix $\mathsf{H}$ appearing in Eq.~\eqref{eq:kernel} does not have any spatial dependence,
then $\bar{\psi}(\bm{x})$ can be readily computed in Fourier space using fast-Fourier transform techniques.
This is the case for isotropically or uniaxially deformed elastomers at constant stiffness, and the numerical minimization
of the free energy in such systems has been described in our previous work~\cite{mannattil2025}.
However, if the entries of $\mathsf{H}$ vary in space, then Eq.~\eqref{eq:blur} is no longer a convolution,
and we need an alternative method to compute $\bar{\psi}(\bm{x})$.

Consider the situation where $\mathsf{H}$ is a diagonal matrix whose entries $\mathsf{H}_{xx} = \mathsf{H}_{yy} = \mathsf{H}_{zz} = h(\bm{x})$
vary in space in addition to the long-wavelength modulus $M_{0}(\bm{x})$.
This is the case, for example, in Fig.~\ref{fig:varstiff}, where both
these quantities vary in space due to a stiffness gradient in the elastomer.
In line with Eq.~\eqref{eq:blur}, we continue to compute $\bar{\psi}(\bm{x})$ as
\begin{equation}
  \bar{\psi}(\bm{x}, t) = \int \dd^{3}\bm{y}\, K_{h(\bm{x})}(\bm{x} - \bm{y})\, \psi(\bm{y}, t),
  \label{eq:variable_h}
\end{equation}
and  $K_{h(\bm{x})}$ is a Gaussian kernel of the form
\begin{equation}
K(\bm{x} - \bm{y}) = \abs{2\pi h^{2}(\bm{x})}^{-3/2}\mathrm{e}^{-\frac{1}{2}\abs{\bm{x} - \bm{y}}^{2}/h^{2}(\bm{x}^{})}.
\end{equation}
As Eq.~\eqref{eq:variable_h} is no longer a convolution, to proceed, we divide the numerical range
of the $h(\bm{x})$ field into $n$ evenly spaced points $(h_{1},
\ldots, h_{n})$ with $h_{1} = \min_{\bm{x}} h(\bm{x})$ and $h_{n} = \max_{\bm{x}} h(\bm{x})$.
We then approximate $\bar{\psi}(\bm{x}, t)$ as a weighted sum
\begin{equation}
  \bar{\psi}(\bm{x}, t) \approx \sum_{i = 1}^{n} c_{i}(\bm{x})\,\bar{\psi}_{h_{i}}(\bm{x}, t),
  \label{eq:variable_h_approx}
\end{equation}
where each $\bar{\psi}_{h_{i}}$, obtained as a Gaussian convolution at a fixed length $h_{i}$, is given by
\begin{equation}
  \bar{\psi}_{h_{i}}(\bm{x}, t) = \int \dd^{3}y\, K_{h_{i}}(\bm{x} - \bm{y})\, \psi(\bm{y}, t).
\end{equation}
We choose the time-independent coefficient fields (the weights) $c_{i}(\bm{x})$
in Eq.~\eqref{eq:variable_h_approx} so that it becomes a linear interpolation
between the different $\bar{\psi}_{h_{i}}$, i.e.,
\begin{equation}
  c_{i}(\bm{x}) = \frac{h_{i + 1} - h(\bm{x})}{h_{i+1}  -  h_{i}}
  \quad\text{and}\quad
  c_{i + 1}(\bm{x}) = \frac{h(\bm{x}) - h_{i}}{h_{i+1} - h_{i}},
\end{equation}
for $h_{i} \leq h(\bm{x}) \leq h_{i + 1}$.
Based on our numerical experiments, we found that choosing $n=10$ interpolation points is a good trade-off between computational speed and accuracy.

Assuming that the local swelling around a point $\bm{x}$ is isotropic, the total elastic free energy
$\mathscr{F}_{\text{el}}$ appearing in Eq.~\eqref{eq:free_total_simple} can be written in real space as
\begin{equation}
  \mathscr{F}_{\text{el}}[\psi] = \frac{1}{2}\int\dd^{3}x\,M_{0}(\bm{x})\left[\bar{\psi}(\bm{x}, t)\right]^{2}.
\end{equation}
Using $\bar{\psi}$ from Eq.~\eqref{eq:variable_h_approx}, we find the chemical potential $\eta^{\text{el}}$ associated with the elastic energy to be
\begin{align}
  \eta^{\text{el}}(\bm{x}, t) &= \frac{\delta \mathscr{F}_{\text{el}}}{\delta \psi}\\
                              &= \sum_{i=1}^{n} \int\dd^{3}y\, K_{h_{i}}(\bm{x} - \bm{y}) M_{0}(\bm{y})\,c_{i}(\bm{y})\,\bar{\psi}(\bm{y}, t)\NN.
\end{align}

To evolve Eq.~\eqref{app:eq:modelb} in time, we first Fourier transform it in space.
Next, we employ Euler discretization for time derivatives with a time step $\delta t$ and a semi-implicit step for the interfacial term, yielding
\begin{equation}
  \psi_{\bm{q}}(t + \delta t) = C^{-1}\left\{A\psi_{\bm{q}}(t) - B\left[b\,\psi_{\bm{q}}^{3}(t) + \eta^{\text{el}}_{\bm{q}}(t)\right]\right\},
  \label{app:eq:modelb_approx}
\end{equation}
where $\psi^{3}_{\bm{q}}$ and $\eta^{\text{el}}_{\bm{q}}$ are the Fourier transforms of $\psi^{3}$ and $\eta^{\text{el}}$, respectively.
Additionally, the time-independent coefficients $A$, $B$, and $C$ are given by
\begin{align}
    A &= 1 - a(T-\Tc)q^{2}\delta t,\NN\\
    B &= q^{2}\delta t,\\
    C &= 1 + \kappa q^{4}\delta t\NN.
\end{align}
To produce Fig.~\ref{fig:varstiff} of the main text, we use a time step of
$\delta t = 5\times10^{-4}$ and evolve Eq.~\eqref{app:eq:modelb_approx}
for $200$ time units in a square domain of side length $10\unit\text{\textmu m}$ and $256 \times 256$ grid points.
Considerably larger time steps $(\delta t \sim 1)$ can be employed by rewriting
Eq.~\eqref{app:eq:modelb_approx} using a linearized splitting scheme when
the length covariance matrix $\mathsf{H}$ is a constant~\cite{mannattil2025}.
All the numerical codes used in this paper are publicly available~\cite{github}.

\bibliography{library,misc}

\end{document}